\documentclass[twocolumn,showpacs,preprintnumbers,draftclsnofoot]{revtex4}
\usepackage{amsmath}
\usepackage{amssymb}
\usepackage{graphicx}
\usepackage{dcolumn}
\usepackage{bm}
\usepackage{amssymb}
\usepackage{graphicx}
\usepackage{dcolumn}
\usepackage{bm}

\begin{document}

\title{ Skin effect and excitation spectral of interacting non-Hermitian system }
\author{Ma Luo\footnote{Corresponding author:luom28@mail.sysu.edu.cn} }
\affiliation{The State Key Laboratory of Optoelectronic Materials and Technologies \\
School of Physics\\
Sun Yat-Sen University, Guangzhou, 510275, P.R. China}

\begin{abstract}

The non-Hermitian Su-Schrieffer-Heeger spinless Fermion model with interacting terms is studied by exact diagonalization. The model is derived from the spin chain with damping Dzyaloshinskii-Moriya interaction. The presence of the interaction induce anomalous spectral function. The skin effect is exhibited by the non-diagonal terms of the Green's function, i.e. the spatial correlation. The interaction modifies the degeneracy between the ground state and the first excited state of the half filled systems, as well as between the ground state of the half and half plus one filled systems. The energy bands of the higher excited states within the topological regime have complicated pattern, depending on the sign and amplitude of the interaction.

\end{abstract}

\maketitle

\section{Introduction}

Topological properties of non-Hermitian systems have been intensively studied recently \cite{Tony16,ShunyuYao18,Kawabata2019,Xufeng2019,RuiChen2019,XiWangLuo2019,ChingHuaLee2019,KazukiYokomizo2019,TianShuDeng2019,FeiSong2019,Stefano2019,Bergholtz2019,TaoLiu2019,Nobuyuki2019,Longhi2019,ZhesenYang19,RebekkaKoch2019,Kazuki2019,Bastian2019,Charles2019}, because of the wide application of the physical models to many reality systems. A typical non-Hermitian model is the one-dimensional  Su-Schrieffer-Heeger (SSH) model. Most of the modeled systems are non-interacting open systems, such as wave systems with gain and loss \cite{Weijian2017,LingLu2014,RamyElGanainy2018}, solid systems of quasi-particle with non-Hermitian self-energy \cite{HuitaoShen2018}, and circuit system with Majorana corner mode \cite{Ezawa2019,LongwenZhou2019,Ezawa2019a}. Comparing to the Hermitian counterpart, the non-Hermitian systems have spatially decaying or amplifying bulk states. The bulk states are described in the generalized Brillouin zone. In the topological regime, the robust zero energy edge modes appear. The non-Bloch bulk-boundary correspondence is determined by the non-Bloch winding number.

For a Fermion system, the wave function of the whole system is antisymmetric under the exchange of two identical particles. Considering this symmetric, the particle density of the non-Hermitian SSH model at half filling is uniform \cite{EunwooLee2019}. The topological properties of the system is proposed to be characterized by the non-Hermitian polarization. Because the analysis is given by exact diagonalization (ED) of the many-body Hamiltonian of the short chain of the SSH model, the algorithm can be directly extended to interacting systems.

In this article, we studied the interacting spinless Fermion SSH model. We firstly show that the model can be derived from a one-dimensional anisotropic Heisenberg spin chain with damping Dzyaloshinskii-Moriya interaction (DMI). The model is solved by ED. The method is reviewed, which gives the energy levels as well as the Green's function. The ground state properties of the model is revealed from the Green's function of the ground state. The diagonal part of the Green's function gives the spectral function, which contain the information of the quasi-particle excitation of the interacting system and local density of state (LDOS). The non-diagonal part of the Green's function gives the correlation between different lattice sites, which exhibit the skin effect. The topological properties of the non-interaction system is featured by the degenerate zero energy mode at the Fermi level. For the many-body system, the degeneracy of the ground state is modified by the interaction. The bands of the energy levels of the higher excited states have complicated pattern depending on the interaction. However, the degeneracy of the energy bands have simple rule that is not dependent on the interaction.

This article is organized as following: In section II, the deduction of the interacting spinless Fermion SSH model from the spin model is given, and the ED method is reviewed. In section III, the numerical result is discussed, including the spectral function, Green's function and excitation spectral. In section IV, the conclusion is given.

\section{theoretical model}

\subsection{From spin model to SSH model}

\begin{figure}[tbp]
\scalebox{0.6}{\includegraphics{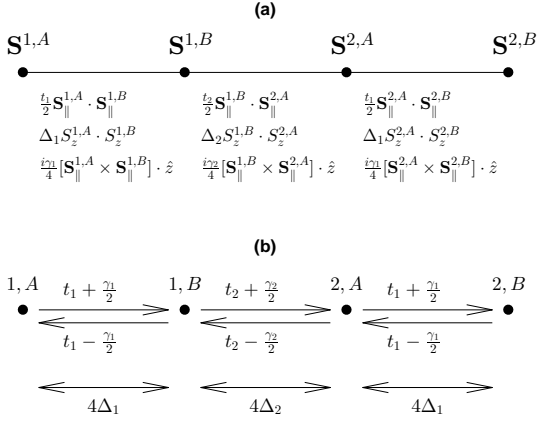}}
\caption{ (a) The Heisenberg spin chain with in-plane coupling being $\frac{t_{1,2}}{2}$, out-of-plane coupling being $\Delta_{1,2}$, damping DMI coupling being $\frac{i\gamma_{1,2}}{4}$. (b) The corresponding non-Hermitian SSH model. The arrows indicate hopping, and the double arrows indicate interaction. \label{fig1}}
\end{figure}

We consider the 1D Heisenberg spin chain as shown in Fig. \ref{fig1}(a). There are two sublattice, designated as A and B, so that each unit cell contains two lattice sites. The number of unit cells in an open chain is designated as $N_{c}$, and the number of lattice sites is $2N_{c}$. The Hamiltonian consists of the coupling between the nearest neighboring spin, which contain three part as $H_{H}=H_{\parallel}+H_{z}+H_{DMI}$. The first part is the in-plane coupling, given as
\begin{equation}
H_{\parallel}=\sum_{k=1}^{N_{c}}{\frac{t_{1}}{2}\mathbf{S}_{\parallel}^{k,A}\cdot\mathbf{S}_{\parallel}^{k,B}}+\sum_{k=1}^{N_{c}-1}{\frac{t_{2}}{2}\mathbf{S}_{\parallel}^{k,B}\cdot\mathbf{S}_{\parallel}^{k+1,A}}
\end{equation}
where $k$ is the index of unit cell. The second part is the out-of-plane coupling, given as
\begin{equation}
H_{z}=\sum_{k=1}^{N_{c}}{\Delta_{1}S_{z}^{k,A}\cdot S_{z}^{k,B}}+\sum_{k=1}^{N_{c}-1}{\Delta_{2}S_{z}^{k,B}\cdot S_{z}^{k+1,A}}
\end{equation}
The third part is the damping DMI, given as
\begin{equation}
H_{\parallel}=\sum_{k=1}^{N_{c}}{\frac{i\gamma_{1}}{4}[\mathbf{S}_{\parallel}^{k,A}\times\mathbf{S}_{\parallel}^{k,B}]\cdot\hat{z}}+\sum_{k=1}^{N_{c}-1}{\frac{i\gamma_{2}}{4}[\mathbf{S}_{\parallel}^{k,B}\times\mathbf{S}_{\parallel}^{k+1,A}]\cdot\hat{z}}
\end{equation}
The difference between the damping DMI and the regular DMI is the additional imaginary factor $i$, which could be interpreted as a ninety degree phase shift at the spin-orbit coupling.

In order to transfer the Heisenberg spin to spinless Fermion model, we define the operator $S^{\pm}_{k,A(B)}=\frac{1}{2}(S_{x}^{k,A(B)}\pm iS_{y}^{k,A(B)})$, and thus $S_{z}^{k,A(B)}=2S^{+}_{k,A(B)}S^{-}_{k,A(B)}-1$. The Hamiltonian becomes
\begin{eqnarray}
H_{H}=\sum_{k=1}^{N_{c}}{(t_{1}+\frac{\gamma_{1}}{2})S^{+}_{k,B}S^{-}_{k,A}+(t_{1}-\frac{\gamma_{1}}{2})S^{+}_{k,A}S^{-}_{k,B}} \nonumber \\+\sum_{k=1}^{N_{c}-1}{(t_{2}+\frac{\gamma_{2}}{2})S^{+}_{k+1,A}S^{-}_{k,B}+(t_{2}-\frac{\gamma_{2}}{2})S^{+}_{k,B}S^{-}_{k+1,A}}\nonumber\\
+\sum_{k=1}^{N_{c}}{\Delta_{1}(2S^{+}_{k,A}S^{-}_{k,A}-1)(2S^{+}_{k,B}S^{-}_{k,B}-1)}\nonumber\\
+\sum_{k=1}^{N_{c}-1}{\Delta_{2}(2S^{+}_{k,B}S^{-}_{k,B}-1)(2S^{+}_{k+1,A}S^{-}_{k+1,A}-1)}
\end{eqnarray}
Applying the Jordan Wigner Transformations to spinless fermion operator $c^{(\dag)}_{k,A(B)}$ at site [k,A(B)], the interacting SSH model is given as
\begin{eqnarray}
&&H_{SSH}=\sum_{k=1}^{N_{c}}{[(t_{1}+\frac{\gamma_{1}}{2})c^{\dag}_{k,B}c_{k,A}+(t_{1}-\frac{\gamma_{1}}{2})c^{\dag}_{k,A}c_{k,B}]} \nonumber \\&&+\sum_{k=1}^{N_{c}-1}{[(t_{2}+\frac{\gamma_{2}}{2})c^{\dag}_{k+1,A}c_{k,B}+(t_{2}-\frac{\gamma_{2}}{2})c^{\dag}_{k,B}c_{k+1,A}]}\nonumber\\
&&+\sum_{k=1}^{N_{c}}{4\Delta_{1}\hat{n}_{k,A}\hat{n}_{k,B}}+\sum_{k=1}^{N_{c}-1}{4\Delta_{2}\hat{n}_{k,B}\hat{n}_{k+1,A}}\nonumber\\
&&+2\Delta_{2}(\hat{n}_{1,A}+\hat{n}_{N_{c},B})+U_{0}
\end{eqnarray}
, where $\hat{n}_{k,A(B)}=c^{\dag}_{k,A(B)}c_{k,A(B)}$ is the particle number operator at each site, $U_{0}=(\Delta_{1}+\Delta_{2})(-2\hat{N}+N_{c})-\Delta_{2}$ is the total energy shift with $\hat{N}=\sum_{k}{\hat{n}_{k,A}+\hat{n}_{k,B}}$ being the total particle number. The model is indicated in Fig. \ref{fig1}(b).

\subsection{Exact diagonalization and Green's function}

The interacting SSH model in open chain can be solved by ED. The basis functions are the Fock states with particle number being $N_{e}$. Because the total site is $2N_{c}$, the half filled system is defined as  $N_{e}=N_{c}$, and the half plus (minus) one filled system is defined as $N_{e}=N_{c}\pm1$. For a given filling factor, only the Fock states with the same particle number is included in the Hilbert space. The matrix form of the Hamiltonian is obtained by operating the SSH Hamiltonian on all Fock states. We define the composite indices $\alpha$ ($\beta$) as $2(k-1)+l$, with $l=1(2)$ for A(B) sublattice. For the hopping between two sites with indices $\alpha$ and $\beta$, the operation of the operator $c^{\dag}_{\alpha}c_{\beta}$ on the Fock state is nonzero if the $\alpha$-th ($\beta$-th) site is unoccupied (occupied). After the operation, the Fock state become another Fock state with the $\alpha$-th ($\beta$-th) site being occupied (unoccupied), and with a sign adjustment $(-1)^{N_{occ}}$. $N_{occ}$ is the number of occupied sites between the $\alpha$-th and $\beta$-th sites. The sign adjustment is due to the anti-commutation relation of the fermion operators. By diagonalization of the matrix form of the Hamiltonian, the eigen-energies and eigen-states are obtained. For a short chain with $N_{c}=5$, full diagonalization of the Hamiltonian can be obtained. For longer chain, iterative solver could find the eigen energies with the smallest real part. Noted that the Lanczos method does not work in this case, because the Hamiltonian is non-Hermitian. In this article, we only studied the short chain with $N_{c}=5$ to exhibit the physical properties of the interacting non-Hermitian systems. For the non-Hermitian matrix $H$, the left and right eigenstates are defined as $H^{\dag}|\Phi_{L,n}\rangle=E_{n}^{*}|\Phi_{L,n}\rangle$ and $H|\Phi_{R,n}\rangle=E_{n}|\Phi_{R,n}\rangle$, respectively, with $E_{n}$ being the n-th eigen-energy, and $|\Phi_{L,n}\rangle$ ($|\Phi_{R,n}\rangle$) being the corresponding left (right) eigen-state.

For the ground state, the Green's function in real space and frequency domain can be calculated as
\begin{eqnarray}
G_{\alpha,\beta}(z)&=&\langle\Phi_{L,1}|c_{\alpha}\frac{1}{z-\mathcal{H}_{+}+Re[E_{1}]}c_{\beta}^{+}|\Phi_{R,1}\rangle \nonumber \\&+&\langle\Phi_{L,1}|c_{\beta}^{+}\frac{1}{z+\mathcal{H}_{-}-Re[E_{1}]}c_{\alpha}|\Phi_{R,1}\rangle
\end{eqnarray}
where $\mathcal{H}_{+(-)}$ is the matrix form of the interacting Hamiltonian in the Hilbert space of half pulse(minus) one filled, $z=\omega+i0$ is the energy with a positive infinitesimal imaginary part. The operation of $c_{\beta}^{(+)}$ on the wave function transfer the Fock state basis into those in the Hilbert space of half pulse(minus) one filled. The LDOS can be obtained by calculating the spectral function from the Green's function as $S(\alpha,\omega)=-\frac{1}{\pi}Im(G_{\alpha,\alpha})$. The particle density in real space is given by integration of $S(\alpha,\omega)$ through the energy up to the Fermi level, $\rho(\alpha)=\int_{\omega=-\infty}^{0}{S[\alpha,\omega]d\omega}$; and the density of states (DOS) is given by the summation of $S(\alpha,\omega)$ through all lattice cite, $S(\omega)=\sum_{\alpha}{S(\alpha,\omega)}$. The non-diagonal matrix of $G(z)$ represent the correlation between two lattice at frequency $\omega$. By integrating the Green's function through the frequency, the zero time delay correlation, $G^{t0}\equiv G(t,t')|_{t'=t}=\int_{\omega}{G(z)d\omega}$, can be obtained.


\section{numerical result and discussion}

We focus the numerical study on the systems with $\gamma_{1}=4/3$, $t_{2}=1$, $\gamma_{2}=0$, and varying $t_{1}$ and $\Delta_{1}=\Delta_{2}$.

\subsection{Density of States}

\begin{figure}[tbp]
\scalebox{0.6}{\includegraphics{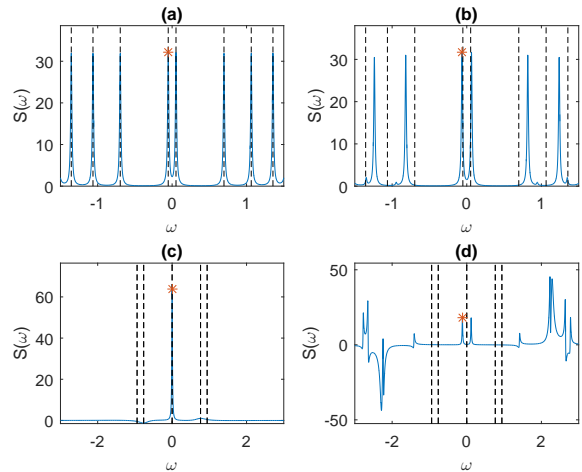}}
\caption{ The spectral functions are plotted as solid lines. The energy level of the corresponding non-interacting system in single particle model is plotted as dashed vertical lines. The systems' parameters are (a) $t_{1}=0.9$, $\Delta_{1}=\Delta_{2}=0$, (b) $t_{1}=0.9$, $\Delta_{1}=\Delta_{2}=0.1$, (c) $t_{1}=0.3$, $\Delta_{1}=\Delta_{2}=0$, (d) $t_{1}=0.3$, $\Delta_{1}=\Delta_{2}=0.7$. $\gamma_{1}=4/3$, $t_{2}=1$ and $\gamma_{2}=0$ for all systems. \label{fig2}}
\end{figure}

\begin{figure}[tbp]
\scalebox{0.6}{\includegraphics{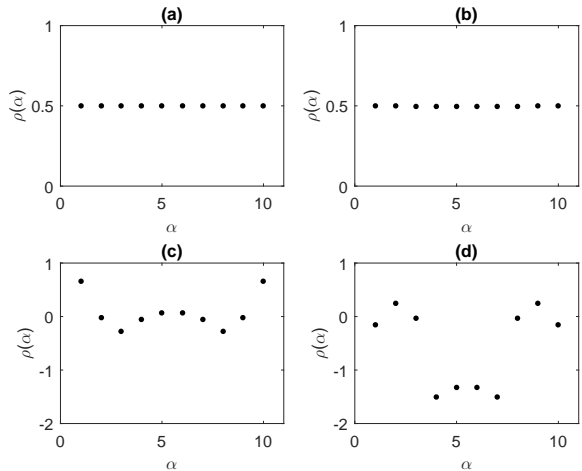}}
\caption{ The particle densities $\rho(\alpha)$ of the same systems as the corresponding sub-figures in Fig. \ref{fig2}. \label{fig3}}
\end{figure}

The DOS of four particular systems are plotted in Fig. \ref{fig2}. The particle densities in real spaces of the same systems are plotted in Fig. \ref{fig3}. The two non-interacting systems in Fig. \ref{fig2}(a) and (c) are topological in the long chain limit, $N_{c}\rightarrow\infty$, because they are within the regime $|t_{1}|<\sqrt{t_{2}^{2}+(\gamma_{1}/2)^2}$ \cite{ShunyuYao18}. For short chain with $N_{c}$ being odd number, the critical value of $|t_{1}|=|\gamma_{1}/2|$ separate the topological phase regime into two parts.

(i) With $|\gamma_{1}/2|<|t_{1}|<\sqrt{t_{2}^{2}+(\gamma_{1}/2)^2}$, the eigen energies are real. The two energy levels near to the Fermi level are not exactly zero, and approach zero as $N_{c}$ increases. The peaks of the DOS match with the eigen-energies of the single particle model, as shown in Fig. \ref{fig2}(a). As the system become interacting, the peaks of the DOS shift, and multiple small peaks emerge, as shown in Fig. \ref{fig2}(b). The particle densities of these system are homogeneous in real space, as shown in Fig. \ref{fig3}(a) and (b).

(ii) With $|t_{1}|<|\gamma_{1}/2|$, two modes with the real part of the energy level being exactly zero appears at the Fermi level. The eigen energies are complex. The DOS is not positive definite, as shown in Fig. \ref{fig2}(c). As the system become interacting, the peak of the DOS at zero energy is split into two peaks; multiple peaks and nadirs emerge in the DOS, as shown in Fig. \ref{fig2}(d). The particle densities of these system are non-homogeneous in real space and are not positive definite, as shown in Fig. \ref{fig3}(c) and (d). The densities are equally localized at both end of the chain, so that the skin effect is not exhibited by the particle density. The anomalous phenomenon of the DOS and particle density of these system is due to the presence of the complex eigen-energies.

\subsection{Spatial correlation}

\begin{figure}[tbp]
\scalebox{0.6}{\includegraphics{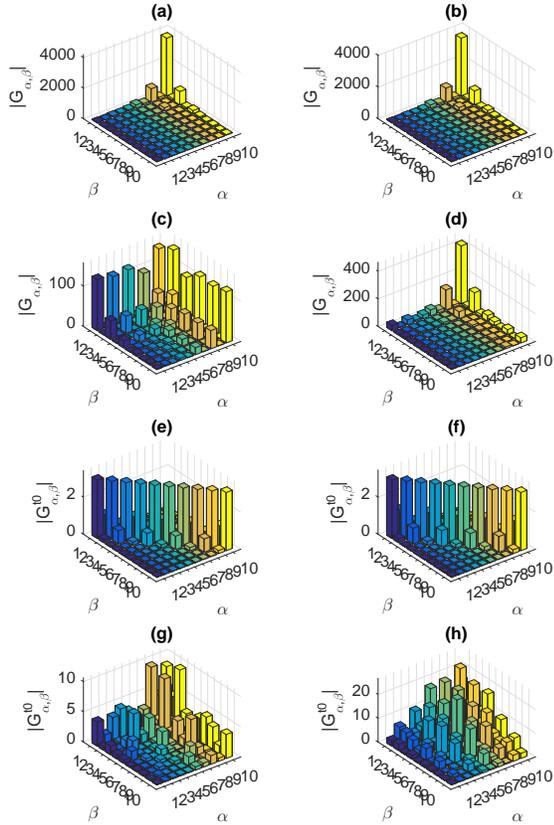}}
\caption{ The amplitude of the Green's function in (a-d), and that of the zero time delay correlation ($G^{t0}$) in (e-h), for the corresponding systems in Fig. \ref{fig2}(a-d). The energy $\omega$ of the figures (a-d) is indicated by the star shape points in the corresponding figures in Fig. \ref{fig2}(a-d). The x and y axis are the composite indices of lattice sites.  \label{fig4}}
\end{figure}

Although the particle density does not exhibit the skin effect, the spatial correlation exhibits the skin effect in many-body systems. The correlation between two lattice sites is given by the non-diagonal matrix elements of the Green's function between them, i.e., $G_{\alpha,\beta}(\omega)$. For the chosen peaks in Fig. \ref{fig2}, the amplitude of the Green's function matrixes are plotted in Fig. \ref{fig4}(a-d). The correlations between the first and the last lattice sites are the most asymmetric, i.e. the amplitude of $G_{1,2N_{c}}$ and $G_{2N_{c},1}$ is largely different. This is the exhibition of the skin effect. For the non-interacting systems, the single particle wave functions of all energy levels are localized near to the 1-st lattice site, and exponentially decaying toward the $N_{c}$-th lattice site. If an excitation is launched at the $\alpha$-th lattice site, the signal will be exponentially decayed (amplified) as it reaches the $N_{c}$-th (1-st) lattice site, due to the exponential form of the excited modes. For the interacting systems, an excitation being launched at the $\alpha$-th lattice site excite quasi-particle that is described by the spectral function. Although the spatial distribution of the quasi-particle is uniform (at least for the case with $|\gamma_{1}/2|<|t_{1}|<\sqrt{t_{2}^{2}+(\gamma_{1}/2)^2}$), the dynamical propagation of the quasi-particle is non-reciprocal. This phenomenon is the most dramatic if the excitation is launched at the open end. The zero time delay correlation $G^{t0}$ of the systems are plotted in Fig. \ref{fig4}(e-h). For the systems with $|\gamma_{1}/2|<|t_{1}|<\sqrt{t_{2}^{2}+(\gamma_{1}/2)^2}$, the amplitude of both $G_{1,2N_{c}}$ and $G_{2N_{c},1}$ are small, as shown in Fig. \ref{fig4}(e) and (f), implying that the skin effect requires finite time delay. For the systems with $|t_{1}|<|\gamma_{1}/2|$, the asymmetric between the amplitude of $G_{1,2N_{c}}$ and $G_{2N_{c},1}$ remain, as shown in Fig. \ref{fig4}(g) and (h), implying that the skin effect could be effective without time delay.

\subsection{Excitation spectral}

\begin{figure}[tbp]
\scalebox{0.6}{\includegraphics{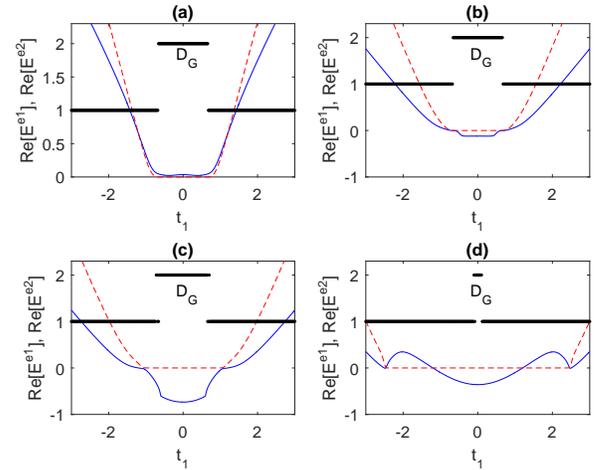}}
\caption{ $E^(e1)$, $E^{(e2)}$, and $D_{G}$ versus $t_{1}$ with $\Delta_{1}=\Delta_{2}=0.5$ in (a), $\Delta_{1}=\Delta_{2}=-0.2$ in (b), $\Delta_{1}=\Delta_{2}=-0.5$ in (c), and $\Delta_{1}=\Delta_{2}=-1$ in (d), are plotted as red dashed lines, blue solid lines, and black dots, respectively. $\gamma_{1}=4/3$, $\gamma_{2}=0$, $t_{2}=1$. \label{fig5}}
\end{figure}

In order to inspect the topological property, we studied the degeneracy between the ground state and the first excited state. There are two ways to define the first excited state: one is the eigenstate of the Hamiltonian of the half filled system with real part of the eigen-energy being the second smallest; the other one is the ground state of the Hamiltonian of the half plus one filled system. The excitation energy of the two types of first excited state from the ground state are designated as $E^{(e1)}$ and $E^{(e2)}$, respectively. For the non-interacting topological system with $|t_{1}|<|\gamma_{1}/2|$, the real part of both $E^{(e1)}$ and $E^{(e2)}$ are zero, i.e. the real part of the ground state is two degree degenerated. Assuming that the system is half filled, one of the two zero energy modes at the Fermi level are occupied. Either adding one particle to the system or moving the particle at one zero energy mode to the other does not change the total energy. In addition to the ground state degeneracy, the two eigen-states coalesce. In the many-body case, the two degenerated ground states have energies and eigen-vectors being complex conjugate to each other. We designated the coalescing between the ground state and the first excited state as $D_{G}$, which is two if the two states coalesce; one otherwise. For the non-interacting system, $D_{G}$ equate two within the range $|t_{1}|<|\gamma_{1}/2|$, and one outside of this range.

In the presence of interaction, the degeneracy of the topological state is partially broken as shown by the numerical result.

(i) In the presence of positive interaction, $E^{(e1)}$ is zero within the range $|t_{1}|<|\gamma_{1}/2|$, and is larger than zero outside of this regime. On the other hand, $E^{(e2)}$ is larger than zero for all $t_{1}$, as shown in Fig. \ref{fig5}(a). For the topological systems, the presence of an addition particle change the energy. As a result, the interaction partially break the degeneracy of the topological state. The interaction does not change $D_{G}$.

(ii) In the presence of negative interaction, $E^{(e1)}$ is zero within the range $|t_{1}|<|\gamma_{1}/2|$. If the interaction is small, $E^{(e1)}$ is non-zeros outside of the range $|t_{1}|<|\gamma_{1}/2|$; $E^{(e2)}$ is negative inside of the range $|t_{1}|<|\gamma_{1}/2|$; $D_{G}$ does not change, as shown in Fig. \ref{fig5}(b). For the topological systems within the regime $|t_{1}|<|\gamma_{1}/2|$, the presence of an addition particle reduce the total energy. If the interaction is larger than certain critical value, the range that $E^{(e1)}$ equates to zero expands; $E^{(e2)}$ is negative inside of this range; the range that $D_{G}$ equates to two slightly expands, as shown in Fig. \ref{fig5}(c). As the interaction further increase, the range that $E^{(e1)}$ equates to zero further expands; $E^{(e2)}$ become positive at the two end of this range; the range that $D_{G}$ equates to two shrink, as shown in Fig. \ref{fig5}(d). At certain value of $t_{1}$ given by the crossing of the two lines in Fig. \ref{fig5}(d), both $E^{(e1)}$ and $E^{(e2)}$ equate to zero, so that the degeneracy of the topological state is regained, but the two degenerated ground states do not coalesce.

\begin{figure}[tbp]
\scalebox{0.6}{\includegraphics{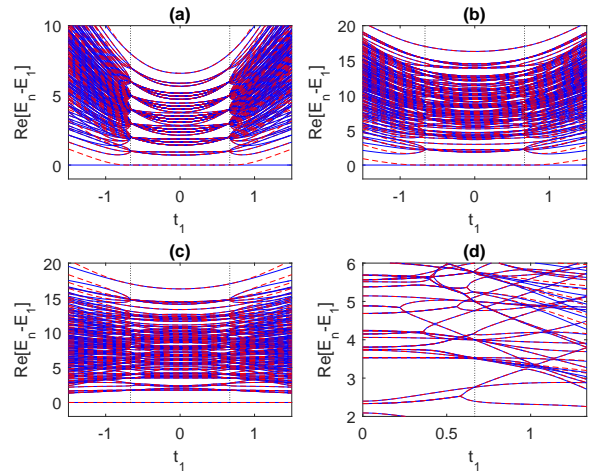}}
\caption{ The real part of the energy levels of all many-body excited states, $E_{n}-E_{1}$, versus $t_{1}$ of the systems with   $\Delta_{1}=\Delta_{2}=0$ in (a), $\Delta_{1}=\Delta_{2}=1$ in (b), and $\Delta_{1}=\Delta_{2}=-1$ in (c). (d) is the zoom in of (c). The energy bands with $n$ being odd (even) are plotted as blue solid (red dashed) lines. The vertical black dotted lines indicate $t_{1}=\pm\gamma/2$.  \label{fig6}}
\end{figure}

The energy levels of all excited state relative to the ground state of the half filled systems are plotted in Fig. \ref{fig6}. For the non-interacting system, the merging points appear at the critical value of $t_{1}=\pm\gamma/2$, as shown in Fig. \ref{fig6}(a), where multiple branches of energy bands merge together. At this point, the single particle bulk band width become zero. For the interacting system with positive and negative interaction, the band structure become more complicated, as shown in Fig. \ref{fig6}(b) and (c), respectively. The zoom in of Fig. \ref{fig6}(c) is plotted in Fig. \ref{fig6}(d), which shows that the merging points of the band structure spread within the range of $|t_{1}|\le\gamma/2$. Although the excited state have complicated band structure, the degeneracy of the excited state have simple structure: (i) Within the range of $|t_{1}|<\gamma/2$, energy levels are second degree degenerated (blue solid lines and red dashed lines overlap in pair in Fig. \ref{fig6}); the ground state and the first excited state could coalesce depending on the system parameters, and all higher pairs of degenerated excited states coalesce. (ii) Outside the range of $|t_{1}|<\gamma/2$, the energy levels might be degenerated accidentally; none of the degenerated pairs of eigen-states coalesce.

\section{conclusion}

In conclusion, the many-body interacting non-Hermitian SSH model in 1D exhibit different features from the single particle non-interacting counterpart. For the topologically trivial phase, the particle density is uniform, and the skin effect is characterized by the spatial correlation. For the topologically non-trivial phase, anomalous features appear, such as negative particle density and zero time delay skin effect. The ground state degeneracy of the topological non-trivial phase is partially broken by the interaction. The excitation spectral have more complicated pattern with many merging points between the energy bands within the topological regime. The exotic phenomenon due to many-body interaction could be original from the exchange correlation physics of the identical Fermions.

\begin{acknowledgments}
This project is supported by the National Natural Science Foundation of China (Grant:
11704419).
\end{acknowledgments}

\clearpage

\end{document}